\documentstyle[prb,aps,graphicx,multicol]{revtex}
\input psfig.sty

\draft

\begin{document}
\title{Study of the ionic Peierls-Hubbard model \\
using density matrix renormalization group methods}
\author{Y. Z. Zhang$^1$, C. Q. Wu$^2$, and H. Q. Lin$^3$}
\address{$^1$Max-Planck-Institut f\"ur Physik komplexer Systeme, \\
N\"othnitzer Stra\ss e 38 01187 Dresden, Germany\\
$^2$Research Center for Theoretical Physics, Fudan University, \\
Shanghai 200433, China\\
$^3$Department of Physics, The Chinese University of Hong Kong, Hong \\
Kong, China}
\date{\today}
\maketitle

\begin{abstract}
Density matrix renormalization group methods are used to investigate the
quantum phase diagram of a one-dimensional half-filled ionic Peierls-Hubbard
model at the antiadiabatic limit where quantum phonon fluctuations are taken
into account partially. We found that two continuous phase transitions
always exist from dimerized spin-gapped (bond-order-wave) state to
band-insulator and undimerized spin-gapless (Mott-insulator) phase while
undimerized spin-gapless phase vanishes at the adiabatic limit. Our results
indicate that quantum phonon fluctuations, electron-electron interaction and
ionic potential combine in the formation of the bond-order wave phase.
\end{abstract}


\begin{multicols}{2}

The response of correlated electrons to lattice distortions in solids has
been extensively studied over the years, due to its important role in
several classes of materials including high-$T_c$ cuprates, colossal
magnetoresistance manganites, conducting polymers and organic
charge-transfer salts. As a good example, the Peierls-Hubbard model, with
on-site Coulomb repulsion and lattice displacement, is a simple yet
nontrivial model that exhibits a rich ground state phase diagram. Strong
correlations lead to the separation of charge and spin excitations \cite
{Emery} while quantum phonon fluctuations can destroy an ordered gapped
state \cite{WuCQ,Bursill,Sengupta2}. Taking them both into consideration is
essential for a full understanding of the nature of these materials.

With the inclusion of additional terms to the Peierls-Hubbard Hamiltonian on
different physics background, various one-dimensional correlated electronic
models were actively studied recently such as the ionic Peierls-Hubbard
model \cite{Nagaosa,Painelli-Girlando,Caprara,Freo} which is defined as
follows 
\begin{eqnarray}
H &=&-\sum_l\left[ t-\alpha \left( u_l-u_{l+1}\right) \right]
B_{l,l+1}+\Delta ~\sum_{l\sigma }(-1)^ln_{l\sigma }  \nonumber \\
&&+U\sum_l\left( n_{l\uparrow }-\frac 12\right) \left( n_{l\downarrow }-%
\frac 12\right)  \nonumber \\
&&\frac 12K\sum_l\left( u_l-u_{l+1}\right) ^2+\frac 1{2M}\sum_lp_l^2,
\label{IPH}
\end{eqnarray}
where $n_{l\sigma }$ is the number operator at site $l$, $\Delta $ is
electrostatic potential of cations, and anions in charge-transfer salts, $%
u_l\left( p_l\right) $ is the displacement (momentum) of the site $l$, $%
\alpha $ and $K$ are the constant for the electron-phonon coupling and
lattice elasticity, $M$ is the mass and the bond-charge density operator $%
B_{l,l+1}$ is 
\begin{equation}
B_{l,l+1}=\sum_\sigma \left( c_{l,\sigma }^{\dagger }c_{l+1,\sigma
}+c_{l+1,\sigma }^{\dagger }c_{l,\sigma }\right) .
\end{equation}
Here we only considered half-filled case.

At $U=0$, model (\ref{IPH}) can be solved exactly at the adiabatic limit. In
Appendix A, we give the detail process and discuss the definitions of the
quantities we used below. Only one phase transition can be found from
dimerized state which is also called Bond-Order-Wave (BOW) state to
Band-Insulator (BI) phase as a function of electron-phonon (e-p) coupling
strength. At $\Delta =0$, it is well-known that no phase transition will
occur even when $U$ goes to infinity as long as there exists e-p coupling at
the adiabatic limit. At $U\neq 0$ and $\Delta \neq 0$, earlier work \cite
{Nagaosa,Painelli-Girlando} and recent work \cite{Caprara,Freo} which
studied the e-p interaction only in the adiabatic limit also concluded that
only one phase transition, i.e., from the BOW to the BI phase, will be
present. In other words, the ionic phase with one electron per site is
always dimerized. Therefore, the phase diagram of this model at the
adiabatic limit is obvious. Only two phases, i.e. BOW and BI phase, can be
detected.

When the lattice distortion is absent, Eq. (\ref{IPH}) represents the Ionic
Hubbard Model (IHM)\cite
{Fabrizio-et,Wilkens-Martin,Torio-et,Zhang-et,Lou-et,Brune-et,Manmana-et},
which was used to describe the neutral-ionic phase transition in mixed-stack
charge transfer crystals \cite{Torrance,Horiuchi-et,Horiuchi-et2,Collet-et}.
As pointed out by Fabrizio, Gogolin and Nersesyan \cite{Fabrizio-et}, there
exists an unusual spontaneously dimerized insulator phase, the BOW phase,
which separates the BI from the Mott insulator (MI) phase \cite{Fabrizio-et}%
. However, in reality, a lattice distortion always exists and it couples to
electronic degrees of freedom strongly in these crystals. Structure changes,
such as volume contraction could be used as an external parameter to drive
the neutral-ionic transition \cite{Horiuchi-et2}. Photoinduced cooperative
phenomena were also observed \cite{Collet-et}. So an important issue to
address is the effect of electron-phonon interactions on the phase diagram.

However, it is well known that results obtained at the adiabatic limit are
unreliable. For the Su-Schrieffer-Heeger (SSH) model, Fradkin and Hirsch 
\cite{Fradkin} pointed out that the low-energy behavior of the system is
actually governed by the antiadiabatic limit $M=0$, rather than the
adiabatic limit $M\rightarrow \infty $. The system at any non-zero frequency
is renormalized to the limit of infinite frequency. For the Holstein model 
\cite{WuCQ} and the spin-Peierls model \cite{Bursill}, more sophisticated
calculations showed that uniform gapless phase exists unless the e-p
coupling is sufficiently large. On the other hand, the system is found
always in the dimerized gapped state at the adiabatic limit. Therefore, in
order to study the effect of e-p interaction truly and understand the whole
phase diagram, it is necessary to investigate the ground state properties of
the system at the antiadiabatic limit.

In the present work, we perform an extensive numerical study of Eq. (\ref
{IPH}) at the antiadiabatic limit using the density matrix renormalization
group (DMRG) \cite{White} technique. We found that, even with very strong
e-p coupling, two continuous phase transitions from BI phase to MI phase are
obtained from this model. One is the spin transition at $U=U_s$ where spin
excitation gap closed, and the other is the charge transition at $U=U_c<U_s$
where the charge excitation gap vanishes. Between these two critical points,
the system is dimerized. In contrast to the adiabatic limit where MI phase
vanishes\cite{Nagaosa,Painelli-Girlando,Caprara,Freo}, this means the ionic
phase can also be undimerized after considering quantum phonon
fluctuations.Our phase diagram is similar to that of IHM\cite
{Torio-et,Wu-suggestion} which is still highly controversial over the years%
\cite
{Fabrizio-et,Wilkens-Martin,Torio-et,Zhang-et,Lou-et,Brune-et,Manmana-et}.
Furthermore, in the region of strong ionic potential, the critical values $%
U_c$ and $U_s$ decrease simultaneously with increasing e-p coupling, while $%
U_s$ will increase with increasing e-p coupling at sufficiently small ionic
potential. The ground state phase diagram is obtained with the use of
finite-size-scaling analysis.

At the antiadiabatic limit $M=0$, an effective interacting fermion model 
\cite{Fradkin} 
\begin{eqnarray}
H &=&-t\sum_lB_{l,l+1}+U\sum_l\left( n_{l\uparrow }-\frac 12\right) \left(
n_{l\downarrow }-\frac 12\right)  \nonumber \\
&&-W\sum_l\left( B_{l,l+1}\right) ^2+\Delta ~\sum_{l\sigma }(-1)^ln_{l\sigma
},  \label{IPH-NA}
\end{eqnarray}
can be obtained where the effective bond-charge attraction ($W\equiv \alpha
^2/2K$) term accounts for the contribution of the phonon quantum
fluctuations. So the Hamiltonian (\ref{IPH-NA}) could be viewed as a 1D e-p
interacting system including both the quantum phonon fluctuations and the
electron correlations.

In this paper, we have applied the finite-size DMRG algorithm with open
boundary conditions to study the Hamiltonian (\ref{IPH-NA}) at half-filling.
This method allows us to probe directly correlation functions and structure
factors associated with the spin density wave (SDW), the charge density wave
(CDW) and the bond order wave in the ground state. Lattices up to $512$
sites were frequently used in our studies. The largest number of states kept
in the calculation was $m=512$ per block. The hopping integral $t$ is set to 
$1$ as the energy unit. The weight of the discarded states was typically
about $10^{-7}-10^{-10}$ depending on whether the system is in its critical
state or not in the final sweep. The convergence tests as functions of
number of states kept were carefully performed. We checked our DMRG
calculations against exact numerical results for noninteracting ($U=W=0$)
chains (up to $512$ sites) and results from exact diagonalization for
interacting ($U\neq 0$, $W\neq 0$) chains (up to $14$ sites). Excellent
agreement was found in both cases. When interactions are turned on, there
exist finite excitation gaps on finite chains, so the accuracies of all
quantities we calculated are no worse than that of the noninteracting case.
Thus, numerical errors in our work could be safely estimated to be smaller
than $10^{-4}$.

\begin{figure}[t b]
\begin{center}
\includegraphics[width=8.5cm]{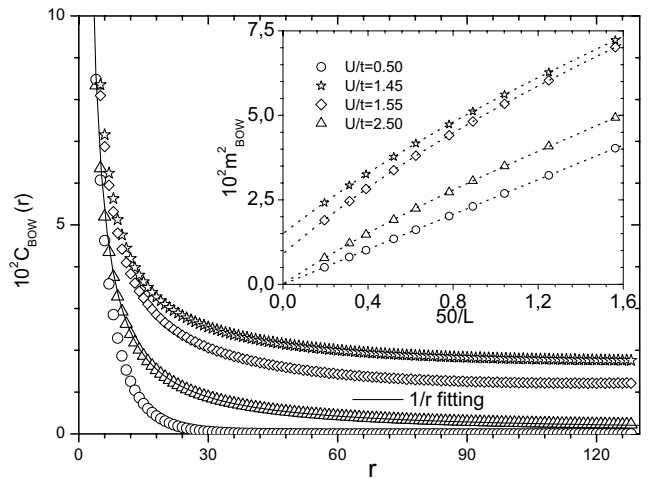}
\end{center}
\begin{minipage}[t]{8.5cm}
\caption{\label{fig:1}
The staggered BOW correlation functions of the ionic
Peierls-Hubbard model for $\Delta /t=0.30$ and $W/t=0.30$. The inset shows
an extrapolation of $m_{BOW}^2\left( L\right) $ with a third-order
polynomial in $1/L$.}
\end{minipage}
\end{figure}
Now let's introduce a way to determine the phase boundaries accurately.
Although parts of this method have already been first used in determining
the phase boundaries of IHM\cite{Zhang-et}, we perform this method more
carefully and systematically in this paper than ever by using various
finite-size analysis. First we defined the staggered BOW correlation
function which is the most direct evidence for the long-range BOW state 
\begin{equation}
C_{BOW}\left( r\right) =\left( -1\right) ^r(\frac 1{L_{av}}%
\sum_l\left\langle B_{l,l+1}B_{l+r,l+r+1}\right\rangle -\overline{B}^2)
\label{CFBOW}
\end{equation}
where $\overline{B}=\frac 1L\sum_l\left\langle B_{l,l+1}\right\rangle $. In
Fig. 1, we show the staggered BOW correlation functions with increasing
Hubbard $U$ at $\Delta /t=0.30$ and $W/t=0.30$. Here we only perform the
average in eq. (\ref{CFBOW}) over 256 sites in the middle of the $512$-site
system, i.e. the sum over $l$ is from $129$ to $256$ and $L_{av}=128$, in
order to further avoid boundary effects. $L$ is the half of the chain length 
$N$ which is located in the central of the system to further eliminate edge
effects. The results indicate that there exist three different phases in
model (\ref{IPH-NA}) since the staggered BOW correlation functions show
three distinct type of behavior as $r$ increases: (i) it decays
exponentially at $U/t=0.50$, indicating that the system has no BOW order;
(ii) it converges to a nonzero constant at $U/t=1.45$ and $U/t=1.55$,
indicating that the system is in the BOW phase with a finite width; (iii) it
decays as $1/r$ at $U/t=2.50$, indicating that the system is in another
phase. The BOW order parameter in the thermodynamic limit

\begin{equation}
\Delta _{BOW}=\lim_{L\rightarrow \infty }\frac 1L\sum_l\left( -1\right)
^{l+1}\left\langle B_{l,l+1}\right\rangle  \label{orderpara}
\end{equation}
can be obtained by fitting $m_{BOW}^2\left( L\right) $ (=$\frac 1L%
\sum_rC_{BOW}(r)$, where the sum over $r$ goes to $L$) with a third-order
polynomial in $1/L$ since $m_{BOW}^2\left( L\right) \rightarrow \Delta
_{BOW}^2$ for $L\rightarrow \infty $. The inset of Fig. 1 shows such
extrapolations. In the following we always take $L_{av}=4$ which is enough
to minimize the oscillations due to the open boundary conditions and sum
over $r$ is only take over central $L$ site where $L$ is half of the chain
length $N$ to further eliminate edge effects. We find that $m_{BOW}^2\left(
L\right) $ approaches zero when $U/t=0.50$ and $2.50$ but remains finite
when $U/t=1.45$ and $U/t=1.55$ which indicate that there exists a finite
region of the BOW phase.

\begin{figure}[t b]
\begin{center}
\includegraphics[width=8.5cm]{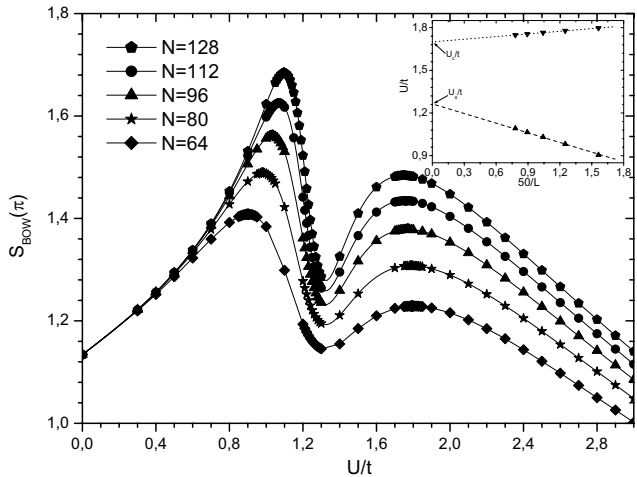}
\end{center}
\begin{minipage}[t]{8.5cm}
\caption{\label{fig:2}
Behavior of $S_{BOW}(\pi)$ across the phase boundary for $%
\Delta/t=0.3$, $W/t=0.30$. The inset shows a linear extrapolation of the
critical values $U_c$ and $U_s$ with $1/L$.}
\end{minipage}
\end{figure}
Next we have studied the nature of BI-BOW and BOW-MI transitions by
calculating the static structure factors corresponding to different phases
to determine the phase boundaries. The first structure factor studied is

\begin{eqnarray}
S_{BOW}(q) &=&\frac 1{L_{av}}\sum_{lr}e^{iqr}(\left\langle
B_{l,l+1}B_{l+r,l+r+1}\right\rangle  \nonumber \\
&&-\left\langle B_{l,l+1}\right\rangle \left\langle
B_{l+r,l+r+1}\right\rangle )~.  \label{connectCF}
\end{eqnarray}
According to Fabrizio et. al.\cite{Fabrizio-et}, phase transitions on the
BI-BOW and BOW-MI phase boundaries are an Ising type and KT type
respectively, the staggered connected correlation function falls off
algebraically as

\begin{equation}
\left( -1\right) ^r(\left\langle B_{l,l+1}B_{l+r,l+r+1}\right\rangle
-\left\langle B_{l,l+1}\right\rangle \left\langle B_{l+r,l+r+1}\right\rangle
)\sim r^{-\eta }
\end{equation}
\begin{figure}[t b]
\begin{center}
\includegraphics[width=8.5cm]{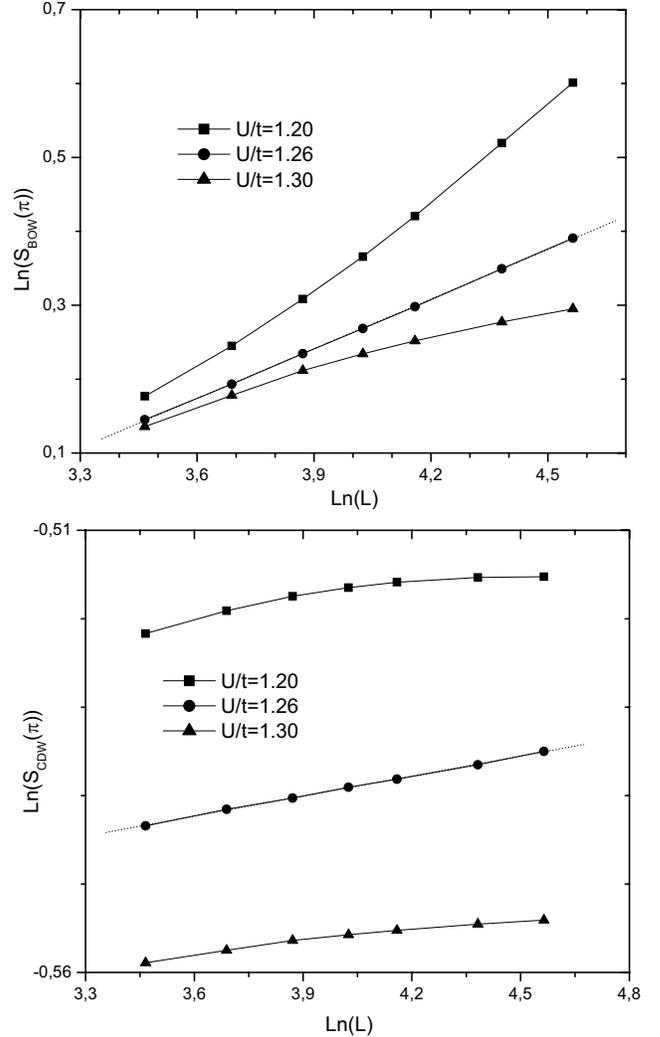}
\end{center}
\begin{minipage}[t]{8.5cm}
\caption{\label{fig:3}
Finite-size analysis for $S_{BOW}(\pi)$ and $S_{CDW}(\pi)$ in the
vicinity of the first phase transition at $U_c$ for $\Delta/t=0.30$, $%
W/t=0.30$.}
\end{minipage}
\end{figure}
Away from phase boundaries, this quantity falls off exponentially. Therefore
the $S_{BOW}\left( \pi \right) $ is expected to diverge at these two
critical points if $\eta \leq 1$ or reach maximums if $\eta >1$ as the
system size goes to infinity. Fig. 2 shows the results of the $S_{BOW}\left(
\pi \right) $ for different system sizes with $\Delta /t=0.30$, $W/t=0.30$
and $0<U/t<3$. As expected, the $S_{BOW}\left( \pi \right) $ peaks twice for
all the different system sizes we calculated. The positions of these two
peaks become closer as the system size is larger. The inset of Fig. 2 shows
linear extrapolations of the positions of these two peaks with $1/L$. We
find that these two peaks will not merge at $L\rightarrow \infty $ which
indicate again that the BOW phase remains finite at thermodynamic limit. In
order to give more convincing evidence, we did another finite-size analysis
in the vicinity of these two phase transitions. Let us start from the first
phase transition at $U=U_c$. Fig. 3(a) presents plots of $\ln [S_{BOW}\left(
\pi \right) ]$ versus $\ln [L]$ for $\Delta /t=0.30$, $W/t=0.30$ and three
different values of $U/t$ around the first critical point. Data points for $%
U/t=1.26$ indeed fall on a straight line, indicating critical scaling for
the BOW fluctuations. At the other two points $U/t=1.20$ and $1.30$, data
points behave nonlinearly due to the exponential decay term. Applying the
same finite-size analysis to the CDW structure factor $S_{CDW}\left( \pi
\right) $, we can also explore the nature of the first phase transition. The
CDW structure factor is defined as 
\begin{equation}
S_{CDW}\left( q\right) =\frac 1{L_{av}}\sum_{lr}e^{iqr}\left( \left\langle
n_ln_{l+r}\right\rangle -\left\langle n_l\right\rangle \left\langle
n_{l+r}\right\rangle \right)
\end{equation}
The linear behavior of $\ln [S_{CDW}\left( \pi \right) ]$ around $U/t=1.26$,
shown in Fig. 3(b), confirms the vanishing of the charge gap at the first
phase transition point.

\begin{figure}[t b]
\begin{center}
\includegraphics[width=8.5cm]{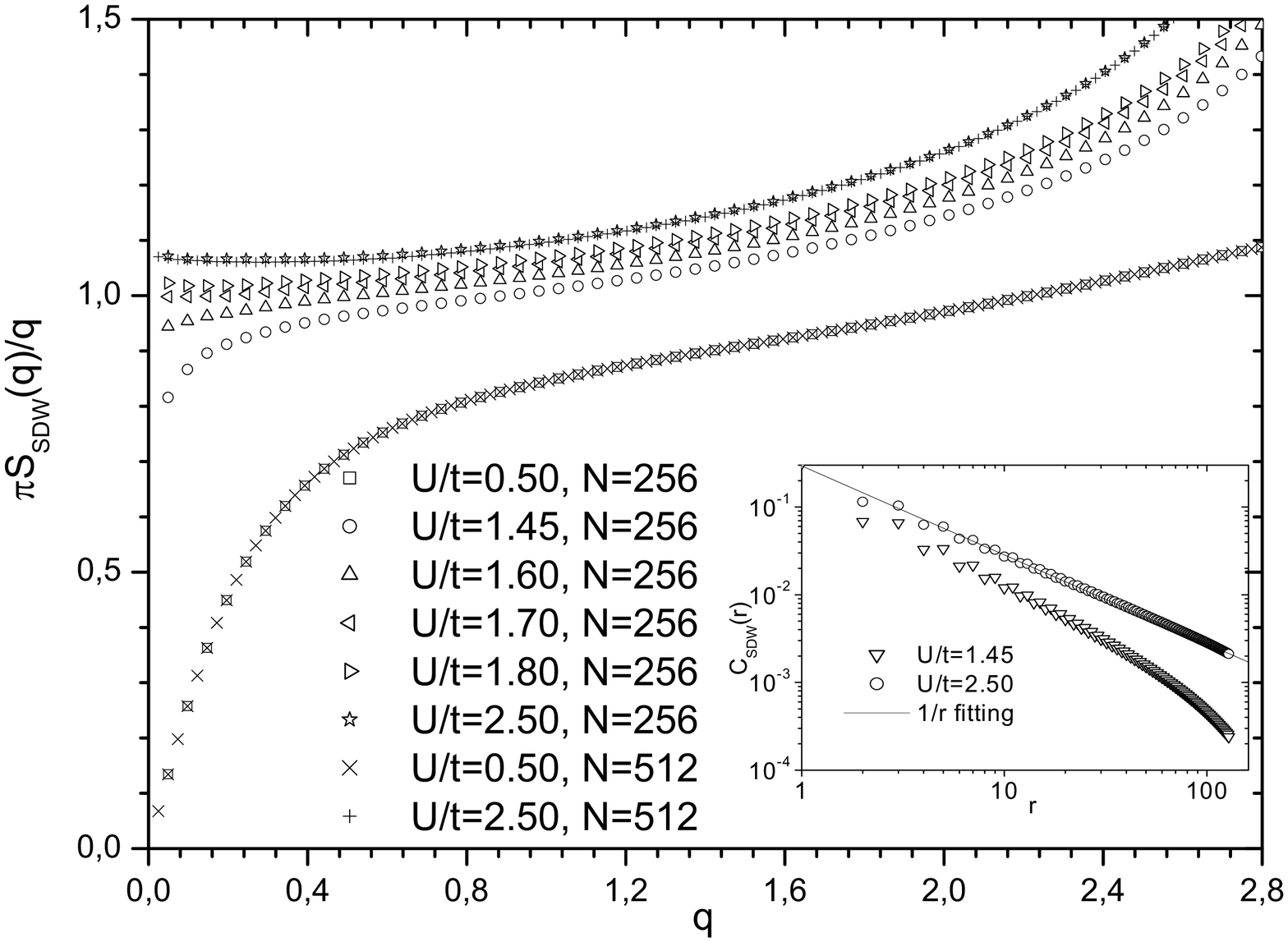}
\end{center}
\begin{minipage}[t]{8.5cm}
\caption{\label{fig:4}
$\pi S_{SDW}(q)/q$ vs $q$ for $\Delta/t=0.30$, $W/t=0.30$ and 6
different values of $U$ across the MI-BOW boundary. }
\end{minipage}
\end{figure}
Next we determine the nature of the second phase transition at $U=U_s$. As
predicted by Fabrizio et. al.\cite{Fabrizio-et}, it is a quantum phase
transition of the KT type. This makes it difficult to determine the phase
boundary directly from the behavior of $S_{BOW}\left( \pi \right) $ and $%
S_{SDW}\left( \pi \right) $ (defined below) due to the finite-size effects.
Instead, we apply an indirect method, used by Sengupta et al.\cite{Sengupta2}%
, to confirm the second phase transition. The SDW structure factor is
defined as

\begin{equation}
S_{SDW}\left( q\right) =\frac 1{L_{av}}\sum_{lr}e^{iqr}\left\langle
s_l^zs_{l+r}^z\right\rangle .
\end{equation}
It is well known \cite{Voit} that if the ground state of a 1D system is
spin-gapless, the spin-spin correlation falls algebraically with exponent
equal to $1$. It has been further shown\cite{Clay} that in the spin-gapless
phase $S_{SDW}\left( q\right) /q\rightarrow 1/\pi $ as $q\rightarrow 0$
whereas in the spin-gapped phase $S_{SDW}\left( q\right) /q\rightarrow 0$.
Even a very small spin gap can be detected in this way, since it is in
practice sufficient to see the $\pi S_{SDW}\left( q\right) /q$ decay below $%
1 $ for small $q$ to conclude that a spin gap must be present. Fig. 4 shows
the behavior of $\pi S_{SDW}\left( q\right) /q$ for $\Delta /t=0.30$, $%
W/t=0.30$ and different values of $U/t$. In the gapless region, logarithmic
corrections \cite{Eggert} make it difficult to observe the approach to $1$
as $q\rightarrow 0$. In analogy with spin systems\cite{Eggert2}, we expect
the leading logarithmic corrections to vanish at the point where spin gap
opens and therefore exactly at the critical point there should be a clear
scaling to $1$. Based on results shown in Fig. 4, we estimate the MI-BOW
boundary to be at $U/t=1.70\pm 0.02$ at $\Delta /t=0.30$, $W/t=0.3$ which is
consistent with the results shown in Fig. 2. The inset of Fig.4 further
provides evidence for the transition from the spin-gapped state, as
identified by the exponential decay of the staggered SDW correlation
function, to the spin-gapless state, as characterized by the $1/r$-decay of
the SDW correlation function.

\begin{equation}
C_{SDW}\left( r\right) =\frac 1L\left( -1\right) ^r\sum_l\left\langle
s_l^zs_{l+r}^z\right\rangle
\end{equation}

\begin{figure}[t b]
\begin{center}
\includegraphics[width=8.5cm]{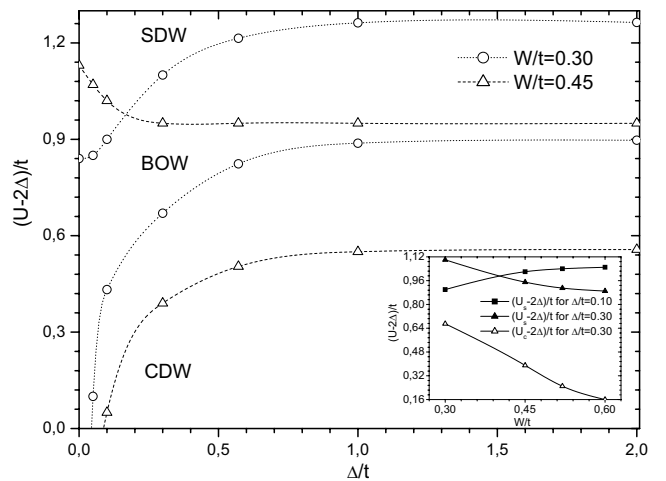}
\end{center}
\begin{minipage}[t]{8.5cm}
\caption{\label{fig:5}
Phase diagram of model (3) for two different e-p couplings. The
inset shows different behavior of $U_s$ for weak ($\Delta/t=0.10$) and
strong ($\Delta/t=0.30$) ionic potentials as a function of e-p coupling W.
While the behavior of $U_c$ as a function of W is always the same for all
ionic potentials. Here we only show the plot for $U_c$ at $\Delta/t=0.30$.}
\end{minipage}
\end{figure}
Finally, we present in Fig. 5 the resulting phase diagram in the $U-\Delta $
plane for two values of $W$. For $\Delta =U=0$, model (3) becomes the $t-W$
model\cite{Zhang} which can be mapped from the SSH model at the
antiadiabatic limit. This model has been studied by the DMRG method\cite
{Zhang} and the renormalization group analysis\cite{Fradkin}. The ground
state is dimerized and the BOW order parameter is nonvanishing as long as $%
W\neq 0$. After switching on the Hubbard $U$ or a finite ionic potential $%
\Delta $, the BOW phase could be destroyed. The model undergoes quantum
phase transitions from the BOW phase either to the MI phase or the BI phase.
The critical value $U$ and $\Delta $ will increase with increasing e-p
coupling $W$. In the weak ionic potential region, such as $\Delta /t=0.1$,
on increasing the e-p coupling $W$, the transition points $U_c$ and $U_s$
move apart and the separation between $U_c$ and $U_s$ becomes significantly
larger. However, for strong ionic potential, such as $\Delta /t=1.0$, both $%
U_c$ and $U_s$ decrease and the width of the BOW phase increases slightly
with increasing e-p coupling. When the ionic potential is intermediate, such
as $\Delta /t=0.3$, $U_c$ and $U_s$ will also decrease simultaneously while
the width of the BOW phase increases significantly. For model (1) at the
adiabatic limit, a spin gap will always be present\cite{Freo}. However, at
the antiadiabatic limit, the transition from BOW to MI phase always occurs
even though the e-p coupling is sufficiently large.

The inset of Fig. 5 shows the asymptotic behavior of the critical points $%
U_c $ and $U_s$ at $\Delta /t=0.10$ and $\Delta /t=0.30$ with increasing e-p
coupling. For weak on-site potential such as $\Delta /t=0.10$, the critical
point $U_s$ becomes larger with increasing e-p coupling while for strong
on-site potential for example $\Delta /t=0.30$, it decrease monotonously
with the increasing e-p coupling. However the behavior of $U_c$ is the same
for all value of on-site potential, it decreases monotonously as a function
of e-p coupling. Here we only show one set of data at $\Delta /t=0.30$.

In conclusion, we have studied 1D half-filled Ionic Peierls-Hubbard model at
the antiadiabatic limit using the DMRG method. The phase diagram is obtained
by investigating correlation functions and structure factors. In contrast to
the adiabatic limit, the transition from the dimerized spin-gapped state to
the spin-gapless state occurs for any value of the ionic potential $\Delta $%
. Compared to the IHM, the BOW phase always exists in the presence of e-p
coupling $W$. The first critical value $U_c$ for the charge gap always
decreases with increasing $W$ while the second critical value $U_s$ for the
spin gap shows different behavior for weak and strong ionic potential.

It is a pleasure to acknowledge helpful discussions with Y. J. Wang, Y. L.
Liu and P. Thalmeier, X. Q. Wang. This work is supported by the Research
Grants Council of the HKSAR, Project No. CUHK 4037/02P. C. Q. Wu also
acknowledges financial support from the National Natural Science Foundation
of China (No. 19725414 and No. 90103034).
\end{multicols}

\appendix

\section{Exact solution to model (\ref{IPH}) at the adiabatic limit in the
noninteracting case}

In this appendix we solve model (\ref{IPH}) exactly at the adiabatic limit
in the noninteracting case. First we define the dimensionless coupling
constant as $\lambda =2\alpha ^2/Kt$ and dimerized order parameter as $%
\delta =\alpha u/t$ where $\left( -1\right) ^lu=u_l-u_{l+1}$. Then model (%
\ref{IPH}) can be rewritten as following 
\begin{equation}
H/t=-\sum_l\left[ 1-\left( -1\right) ^l\delta \right]
B_{l,l+1}+D~\sum_{l\sigma }(-1)^ln_{l\sigma }+\frac 1\lambda \sum_l\delta ^2
\label{IPHNI}
\end{equation}
where $D=\Delta /t$. Using Fourier transformation and unitary
transformation, the original operator $c_{l\sigma }^{+}$ can be expressed as 
\begin{equation}
c_{l\sigma }^{+}=\frac 1{\sqrt{N}}\sum_ke^{ikl}\left( \alpha _k^{*}\left(
l\right) a_{k\sigma }^{+}+\beta _k^{*}\left( l\right) b_{k\sigma }^{+}\right)
\end{equation}
where 
\begin{equation}
\alpha _k\left( l\right) =u_k+\left( -1\right) ^lv_k,\beta _k\left( l\right)
=-v_k^{*}+\left( -1\right) ^lu_k
\end{equation}
and $k\in (-\frac \pi 2,\frac \pi 2]$. Here 
\begin{equation}
u_k=\frac{\sqrt{E_k-2\cos k}}{\sqrt{2E_k}},v_k=\frac{-2i\delta \sin k}{\sqrt{%
2E_k\left( E_k-2\cos k\right) }}
\end{equation}
where 
\begin{equation}
E_k=\sqrt{4\cos ^2k+4\delta ^2\sin ^2k+D^2}.
\end{equation}
Then the Hamiltonian (\ref{IPHNI}) can be diagonalized as 
\begin{equation}
H/t=\sum_{k\sigma }E_k\left( a_{k\sigma }^{+}a_{k\sigma }-b_{k\sigma
}^{+}b_{k\sigma }\right) .
\end{equation}
The quantities related to the BOW\ phase can be expressed as following 
\begin{equation}
\left\langle B_{l+r,l+r+1}\right\rangle =4\left( A_1-\left( -1\right)
^{l+r}B_1\right) .
\end{equation}
here $B_1$ is the order parameter, $A_1$ is the average bond length and 
\begin{eqnarray}
\left\langle B_{l+r,l+r+1}B_{l,l+1}\right\rangle &=&\left\langle
B_{l+r,l+r+1}\right\rangle \left\langle B_{l,l+1}\right\rangle +  \nonumber
\\
&&\left\{ 
\begin{array}{c}
4\left[ C_r^2-\left( A_{r+1}-\left( -1\right) ^lB_{r+1}\right) \left(
A_{r-1}+\left( -1\right) ^lB_{r-1}\right) \right] ,r=even \\ 
4\left[ C_{r+1}C_{r-1}-\left( A_r^2-B_r^2\right) \right] ,r=odd
\end{array}
\right\} .
\end{eqnarray}
where 
\begin{equation}
A_r=\frac 1\pi \int_0^{\frac \pi 2}dk\frac{2\cos kr\cos k}{E_k}.
\end{equation}
\begin{equation}
B_r=\frac \delta \pi \int_0^{\frac \pi 2}dk\frac{2\sin kr\sin k}{E_k}.
\end{equation}
\begin{equation}
C_r=\frac D\pi \int_0^{\frac \pi 2}dk\frac{\cos k}{E_k}.
\end{equation}
Then we find that the definition (\ref{CFBOW}), (\ref{orderpara}), and (\ref
{connectCF}) in our paper is usefull. Since $\overline{B}=\frac 1L%
\sum_l\left\langle B_{l,l+1}\right\rangle =4A_1$, we can obtain 
\begin{eqnarray}
C_{BOW}(r) &=&16B_1^2+  \nonumber \\
&&\left\{ 
\begin{array}{c}
4\left[ C_r^2-\left( A_{r+1}A_{r-1}+B_{r+1}B_{r-1}\right) \right] ,r=even \\ 
4\left[ \left( A_r^2-B_r^2\right) -C_{r+1}C_{r-1}\right] ,r=odd
\end{array}
\right\} .  \label{CFAPP}
\end{eqnarray}
Obviously, if the system is in the BOW phase, $C_{BOW}(r)\rightarrow 16B_1^2$
remains constant at large $r$ otherwise $C_{BOW}(r)\rightarrow 0$ in the CDW
phase since second line in eq. (\ref{CFAPP}) is exponential decay with the
distance $r$ in both phases. The BOW order parameter defined in our paper
can be obtained as 
\begin{equation}
\Delta _{BOW}=\lim_{L\rightarrow \infty }\frac 1L\sum_l\left( -1\right)
^{l+1}\left\langle B_{l,l+1}\right\rangle =4B_1.
\end{equation}
Meanwhile 
\begin{equation}
m_{BOW}^2\left( \infty \right) =\lim_{L\rightarrow \infty }\frac 1L%
\sum_rC_{BOW}(r)=16B_1^2  \label{OPAPP}
\end{equation}
since the second line in eq. (\ref{CFAPP}) is exponential decay with the
distance $r$. Here we should mention that on the phase boundary the second
line in eq. (\ref{CFAPP}) is power-law decay. Nevertheless eq. (\ref{OPAPP})
is still valid. The staggered BOW structure factor is 
\begin{equation}
S_{BOW}(\pi )=\left\{ 
\begin{array}{c}
4\left[ C_r^2-\left( A_{r+1}A_{r-1}+B_{r+1}B_{r-1}\right) \right] ,r=even \\ 
4\left[ \left( A_r^2-B_r^2\right) -C_{r+1}C_{r-1}\right] ,r=odd
\end{array}
\right\} .
\end{equation}
As we mentioned above, exactly on the phase boundary it is power-law decay
otherwise exponential decay. Finally we give the self-consistent equation
which can be used to determine the phase diagram in the $\Delta -\lambda $
plane 
\begin{equation}
1=\frac{2\lambda }\pi \int_0^{\frac \pi 2}dk\frac{2\sin kr\sin k}{E_k}.
\end{equation}

\begin{multicols}{2}

\end{multicols}

\begin{references}
\bibitem{Wu-suggestion}  For the IHM, the region of BOW phase is so small
that it was argued for its existence in the literature. With the bond charge
attraction term, the BOW region is much larger and we can distinguish the
two phase transitions clearly. We regard this as an indirect confirmation of
the BOW phase in the IHM.
\end{references}
\end{document}